\documentclass{article}

\usepackage{arxiv}

\usepackage[utf8]{inputenc} % allow utf-8 input
\usepackage[T1]{fontenc}    % use 8-bit T1 fonts
\usepackage{hyperref}       % hyperlinks
\usepackage{url}            % simple URL typesetting
\usepackage{booktabs}       % professional-quality tables
\usepackage{amsfonts}       % blackboard math symbols
\usepackage{nicefrac}       % compact symbols for 1/2, etc.
\usepackage{microtype}      % microtypography
\usepackage{lipsum}
\usepackage{graphicx}
\graphicspath{ {./images/} }

\usepackage{tikz}
\usetikzlibrary{arrows}
\usepackage[ruled,vlined]{algorithm2e}
\usepackage{graphics}
\usepackage[first=0,last=9]{lcg}

\usepackage{subcaption}
\usepackage{multirow}
\usepackage{enumitem}

\usepackage{amsmath}

\title{SUBMASSIVE: Resolving Subclass Cycles in Very Large Knowledge Graphs}

\author{
 Shuai Wang\\
  Department of Computer Science,\\ Vrije University Amsterdam, \\1081HV Amsterdam, The Netherlands\\
  \texttt{shuai.wang@vu.nl} \\
   \And
Peter Bloem \\
  Department of Computer Science, \\Vrije University Amsterdam, \\1081HV Amsterdam, The Netherlands\\
  \texttt{p.bloem@vu.nl} \\
  \And
 Joe Raad  \\
LISN, University of Paris-Saclay,\\ 
Orsay, France\\ 
  \texttt{joe.raad@lisn.fr} \\
 \And
  Frank van Harmelen\\
 Department of Computer Science,\\ Vrije University Amsterdam,\\ 1081HV Amsterdam, The Netherlands\\
  \texttt{frank.van.harmelen@vu.nl} \\
}

\begin{document}
\maketitle
\begin{abstract}
Large knowledge graphs capture information of a large number of entities and their relations. Among the many relations they capture, class subsumption assertions are usually present and expressed using the \texttt{rdfs:subClassOf} construct. From our examination, publicly available knowledge graphs contain many potentially erroneous cyclic subclass relations, a problem that can be exacerbated when different knowledge graphs are integrated as Linked Open Data. In this paper, we present an automatic approach for resolving such cycles at scale using automated reasoning by encoding the problem of cycle-resolving to a MAXSAT solver. The approach is tested on the LOD-a-lot dataset, and compared against a semi-automatic version of our algorithm. We show how the number of removed triples is a trade-off against the efficiency of the algorithm.\footnote{The paper has been presented at the 2020 workshop on Large Scale RDF Analytics (LASCAR), a workshop co-located with the 
Extended Semantic Web Conference (ESWC). The code, raw data, related intermediate results, and plotting scripts can be found on GitHub at \url{https://github.com/MaestroGraph/SUBMASSIVE} and Zenodo with DOI \texttt{10.5281/zenodo.14535281}. The resulting refined cycle-free subclass and subproperty subsumption relations are available on Zenodo with DOI \texttt{10.5281/zenodo.3693802}.}
\end{abstract}

% keywords can be removed

\section{Introduction}
\label{sec:Introduction}

% Introduction to knowledge graph and its refinement. 
% Knowledge graphs play an increasingly important role for tasks such as information retrieval, natural language understanding, and decision making in systems of various domains, etc. 
% Common knowledge graphs combine the information edited by the crowd, extracted from unstructured/semi-structured knowledge, or harvested from the web. 
% LOD-a-lot is a queryable dump of the LOD cloud that integrates several hundreds of thousands of datasets with over 28 billion triples. Due to the way LOD-a-lot and its included datasets are constructed and integrated, it is hard to ensure the correctness, which is an important issue for the quality of systems using it. The process of verifying and justifying the entries is called knowledge graph refinement. 

%  Some previous work showed how owl:sameAs can be problematic \cite{joe} in very large scale knowledge graphs. In this work \footnote{The source code and data are available on \url{www.submassive.cc}.}, 

% Motivation 

% Relations between classes are arguably the most fundamental information in knowledge graphs. Such information is captured as class axioms in the form of triples with predicates such as rdfs:subClassOf, owl:equivalentClass and owl:sameAs. 

 Among the many relations knowledge graphs capture, subsumption relations on classes are represented as triples of axioms using \texttt{rdfs:subClassOf}. Ideally, such triples form a structure of hierarchy, a.k.a. a directed acyclic graph (DAG), if not considering reflexive relations. From our research, not all knowledge graphs have such an acyclic class taxonomy. In the case of very large knowledge graphs, integrating information from different sources, the cyclic relations from integrated small knowledge graphs are inherited and there is potentially erroneous cyclic information across different domains. Only when given a cycle-free subclass hierarchy, can we obtain a reliable subclass transitive closure of an entity. It is required for the evaluation of the accuracy of transitive relations regarding different knowledge graph embeddings. 

This paper focuses on resolving potentially erroneous cyclic relations in the aim of obtaining a reliable cycle-free hierarchy with a minimum number of removed relations. In Section \ref{sec:Introduction}, we first study some properties of cyclic subclass assertions (Section \ref{data}) and present related work (Section \ref{related}). Section \ref{method} describes our algorithm for removing potentially erroneous cycles, followed by its implementation details. Finally we evaluate our system and present results in Section \ref{eva} followed by a discussion and our plan for future work in Section \ref{discuss}.

% Focus of the paper
% This paper focus on the information on classes. More specifically, the rdfs:subClassOf \footnote{rdfs:subClassOf is an abbreviation of \url{http://www.w3.org/2000/01/rdf-schema#subClassOf}} relation.  The goal of our work is to generate a reliable hierarchical structure of the very large knowledge graph LOD-a-lot \footnote{\url{http://lod-a-lot.lod.labs.vu.nl/}}.

% It is important to have this subclass hierarchical class structure provides richer semantics on entities' type which can infer other properties. The more specific an entity's type is, the more certain we are about its properties. <importance of this information> However, cyclic subclass statements are barriers for methods.
% NEEDS A FORMAL INTRODUCTION TO KG. \\

% \subsection{Preliminaries}
% \label{data}

\subsection{Preliminaries}
\label{data}
% A formal introduction to knowledge graph
A knowledge graph (KG) $G$ consists of triples of relational information on entities. Such triples are in the form of $(s, p, o)$, where $s$ is an entity called the \emph{subject}, $p$ is a predicate, and $o$ is an entity called the \emph{object}. A relation such as class $A$ is a subclass of class $B$ ($A \sqsubseteq B$) is encoded as (A, \texttt{rdfs:subClassOf}, B). From a graph theory point of view, these triples form a directed multigraph with self-loops (a.k.a. reflexive relations). When studying those triples where $p$ is restricted to rdfs:subClassOf, the subgraph $G_\text{cat}$ is solely about the  subsumption relations on classes. Cyclic class relations come in the form $A \sqsubseteq A$ (a reflexive relation; self-loop), or $A \sqsubseteq B$ and $B \sqsubseteq A$ (size two), or more generally $A_1 \sqsubseteq A_2 \sqsubseteq \ldots \sqsubseteq A_N \sqsubseteq A_1$ (size N).

% explain the properties of the cycles
Such cycles can be harmless. For example, reflexive cycles are simply tautologies (always true) and are therefore redundant. There are four possible inclusion relations between two classes $A$ and $B$: $A \sqsubseteq B$, $B \sqsubseteq A$, $A \equiv B$ ($A \sqsubseteq B$ and $B \sqsubseteq A$), or none of them (including the case of unknown). Due to the transitivity, cycles are only correct when all classes in the cycle are equivalent, otherwise they represent a source of error.

% To avoid confusion, I add a small paragraph about terminology.
In this paper, we use a hybrid approach of logical and network/graph theory methods. To avoid confusion, we clarify the terminology: we use \textit{relation} interchangeably with edge and triple; we let \textit{class} to be the same as \textit{node} and \textit{concept}; we use \textit{graph} for \textit{knowledge graph} or \textit{knowledge base}. Similarly, we use \textit{cycle} in reference to \textit{cyclic subclass assertions}.

\subsection{Related Work}

\label{related}

% This work falls into the category of large-scale knowledge graph refinement.
According to \cite{refinement}, knowledge graph refinement methods can be divided into two main categories: \textit{completing} the knowledge graph with missing knowledge, and \textit{identifying wrongly} asserted information. This work falls into the latter category of approaches, as we attempt to resolve cyclic subclass relations by removing potentially incorrect relations.

% This work falls into the category of large-scale knowledge graph refinement.
% According to \cite{refinement}, such knowledge graph refinement methods can be divided into two main categories, where a distinction is made between approaches aiming at \textit{completing} the knowledge graph with missing knowledge, and ones aiming at \textit{identifying wrongly} asserted information. This work falls into the latter category of approaches, as we attempt to resolve cyclic subclass relations by removing potentially incorrect relations.

To the best of our knowledge, this is the first work that specifically focuses on detecting incorrect \texttt{rdfs:subClassOf} relations in large knowledge graphs. In contrast with other types of relations, finding such erroneous class subsumption assertions has attracted less attention, possibly due to the fact that the creation of such assertions is rarely automated, and hence are less prone to error. Focusing on other types of relations, Ma el al. \cite{ma2014learning} proposed a method for detecting wrong type assertions by relying on disjointness axioms which they lean using inductive logic programming. Paulheim and Bizer \cite{paulheim2014improving} proposed a statistical method for finding erroneous statements for each type of relation by identifying edges whose subject and object type strongly deviate from the characteristic distribution. The largest amount of work on knowledge graph refinement have focused on detecting erroneous equivalence and identity relations. In this family of approaches, various kinds of information have been exploited to identify erroneous statements, such as the description related to the linked resources \cite{paulheim2014identifying,cuzzola2015filtering}, domain knowledge that is included in the ontology or obtained from experts \cite{hogan2012scalable,papaleo2014logical}, and different network metrics \cite{gueret2012assessing,raad2018detecting}.

\section{Algorithm}
\label{method}

The algorithm consists of two parts: data pre-processing as in Section \ref{pre} and the automated refinement as in Section \ref{auto}. Implementation details are included in Section \ref{implementation}.

\subsection{Data Pre-processing}
\label{pre}
The subgraph $G_\text{cat}$ is a collection of all the triples in the form (s, \texttt{rdfs:subClassOf}, o). For the sake of efficiency, those classes without any subclasses (leaf nodes) are temporally removed, since by definition they can not be part of any cycle. We further remove all reflexive relations since they form trivial self-cycles. Both of these operations can be done in  linear time regarding the number of nodes. 

% Next we take into account the equivalent classes and that of the same concept across domain. An rdfs:subClassOf relation between s and o is removed if  (s, owl:equivalentClass, o) or (s, owl:sameAs, o) is in the graph, since the subclass relation is then subsumed by the equivalentClass or sameAs relation. Assuming a properly indexed triple store, this operation is again linear in the size of the graph.

In addition, we make use of existing equivalence and identity relations to make direct decisions regarding certain subsumption relations. Specifically, we remove all \texttt{rdfs:subClassOf} statements between $s$ and $o$ (as \textit{unnecessary} relations) when there is an explicit assertion of \texttt{owl:equivalentClass} or the transitive closure of \texttt{owl:sameAs}.

\subsection{Automated Cycle-resolving}
\label{auto}
We resolve the cycles iteratively from local neighbourhoods to cross-domain cycles and eventually ensure the entire knowledge graph is cycle-free. This implies that the amount of edges we remove is not minimal but an approximation to it. In each iteration, there are two steps: obtain a small subgraph within which we detect the cycles, which we then encode in a MAXSAT solver which makes decision about which relations to remove. We repeat this process until there is no cycle in the entire graph, or until processing time runs out. This gives us an anytime algorithm which gives us increasingly better approximations of a cycle-free graph. Algorithm~\ref{algorithm:refinement} provides an overview of the method in pseudocode. \\

\begin{algorithm}[]
\SetAlgoLined
\KwResult{A refined knowledge graph with no cyclic subclass assertions}
 initialization: retrieve subclass subsumption subgraph $G_\text{cat}$ from $G$\;
 pre-processing $G_\text{cat}$\;
 \While{$G_\text{cat}$ is not cycle free and not timeout}{
  obtain a set of nodes $N$ with soft upperbound $B$ on size from $G_\text{cat}$\;
%   \eIf{condition}{
    obtain neighbourhood subgraph $G[N]$ corresponding to $N$\;
    generate \textit{simple cycles} $SC$ from $G[N]$\;
   encode $SC$ to a MAXSAT solver and obtain a model $m$\;
   decode the results from $m$ and remove $E'$ accordingly and get $G_\text{cat}'$.
 }
 export all the removed edges $E'$\;
 return $G_\text{cat}'$
 \caption{Refinement of the Subsumption Relations on Classes}
 \label{algorithm:refinement}
\end{algorithm}

\subsubsection{Retrieve Local Simple Cycles.}
Bounded by an upper-bound $B$, we collect a set of nodes by retrieving cycles in $G_\text{cat}$ (see Section \ref{implementation} for implementation details). From these nodes $N$, we obtain the corresponding neighbourhood $G[N]$. Next, we retrieve all the simple cycles within this subgraph.

For a graph, a \textit{simple cycle}, or an \textit{elementary circuit}, is a closed path where no node appears twice except that the first and last node are the same.  The time complexity for a function to obtain all the simple cycles is $O((n+e)(c+1))$ for $n$ nodes (classes), $e$ relations (edges) and $c$ elementary circuits \cite{simple}.

\begin{center}
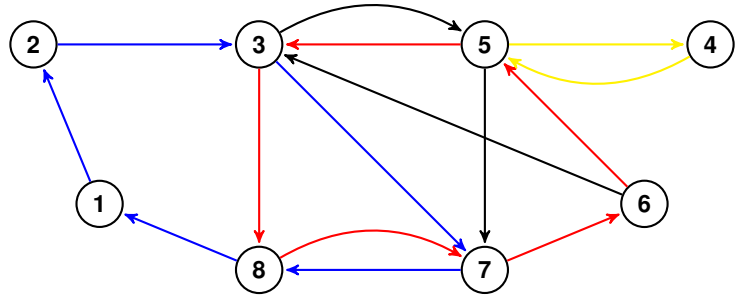

\begin{tikzpicture}[->,>=stealth',shorten >=1pt,auto,node distance=3cm,
                    thick,scale=0.55,main node/.style={circle,draw,font=\sffamily\small\bfseries}]

  \node[main node] (2) {2};
    \node[main node] (3) [right of=2] {3};
  \node[main node] (1) [below left of=3] {1};
  \node[main node] (8) [below of=3] {8};
 \node[main node] (5) [right of=3] {5};
  \node[main node] (7) [below  of=5] {7};
  \node[main node] (6) [below right of=5] {6};
   \node[main node] (4) [right of=5] {4};

  \path[every node/.style={font=\sffamily\small}]
    (1) edge[blue, thick] node [left] {} (2)
    (2) edge [blue, thick] node[left] {} (3)
    (3) edge [red, thick] node[left] {} (8)
    (3) edge [blue, thick] node[left] {} (7)
    (8) edge [blue, thick] node[left] {} (1)
     (8) edge [red, thick, bend left] node[left] {} (7)
     (3) edge [black, thick, bend left] node[left] {} (5)
      (6) edge [red, thick] node[left] {} (5)
      (5) edge [red, thick] node[left] {} (3)
      (7) edge [blue, thick] node[left] {} (8)
      (7) edge [red, thick] node[left] {} (6)
      (6) edge [black, thick] node[left] {} (3)
      (5) edge [black, thick] node[left] {} (7)
      (5) edge [yellow, thick] node[left] {} (4)
        (4) edge [yellow, bend left] node[left] {} (5);
\end{tikzpicture}
	\captionof{figure}{An example graph}
	\label{fig:exampegraph}
\end{center}

Consider the local neighbourhood in Figure \ref{fig:exampegraph}, over nodes $\{1, 2, \ldots, 8\}$, with asserted cycles indicated in blue, yellow and red. 
 %FvH The following is an example graph retrieved from the nodes in the cycles $C$: $1 \rightarrow 2 \rightarrow 3 \rightarrow 7 \rightarrow 8 \rightarrow 1$ (blue), $4 \rightarrow 5 \rightarrow 4$ (yellow) and $5 \rightarrow 3 \rightarrow 8 \rightarrow 7 \rightarrow 6 \rightarrow 5$ (red). Therefore the nodes are $N = \{1, 2, \ldots, 8\}$ and the corresponding graph is as below. 
 The simple cycles are 
$1  \rightarrow
2  \rightarrow
3  \rightarrow
5  \rightarrow
7  \rightarrow
8 \rightarrow 1$, 
$1  \rightarrow
2  \rightarrow
3  \rightarrow
8 \rightarrow 1$,
$1  \rightarrow
2  \rightarrow
3  \rightarrow
7  \rightarrow
8 \rightarrow 1$,
$3  \rightarrow
5  \rightarrow
7  \rightarrow
6 \rightarrow 3$,
$3  \rightarrow 5 \rightarrow 3$,
$3  \rightarrow
8  \rightarrow
7  \rightarrow
6 \rightarrow 3$,
$3  \rightarrow
8  \rightarrow
7  \rightarrow
6  \rightarrow
5 \rightarrow 3$,
$3  \rightarrow
7  \rightarrow
6 \rightarrow 3$,
$3  \rightarrow
7  \rightarrow
6  \rightarrow
5 \rightarrow 3$,
$4  \rightarrow
5 \rightarrow 4$,
$8  \rightarrow
7 \rightarrow 8$, and 
$5  \rightarrow
7  \rightarrow
6 \rightarrow 5$. A non-simple cycle is $5 \rightarrow 7 \rightarrow 6 \rightarrow 5 \rightarrow 3 \rightarrow 8 \rightarrow 7 \rightarrow 5$.

 It is obvious that if all simple cycles in a graph are resolved, there will be no cycle in the graph anymore.\footnote{By \emph{resolving a cycle} we mean that at least one of its edges is removed.}  We therefore list all the simple cycles $SC$ of the corresponding subgraph $G[N]$. Next, we employ a MAXSAT solver to find the smallest number of edges that can be removed to resolve all cycles. 

% % TODO
% With all the simple cycles $SC$, we can encode to a MAXSAT solver to resolve all the cycles. 

% This provides a better way to handle combinatorial explosion than enumerating all the cycles. Despite its speed, the negative side of this approach is that we will not guarantee that the total amount of edges we remove is a global minimum. Moreover, there is a trade off with speed and we set a upper bound of $|N|$ to be 50.

\subsubsection{Resolving Simple Cycles Using MAXSAT.}
In each iteration (i.e. for each local neighbourhood, as in Figure~\ref{fig:exampegraph}), we obtain a set of simple cycles $SC$ from the subgraph $G[N]$ as described above. Next, in order to remove the minimum amount of edges to break these cycles, we employ a MAXSAT solver and encode all the cycles to it. 

% \begin{figure}[!ht]
%     \centering
%   \begin{tikzpicture}[->,>=stealth',shorten >=1pt,thick]
% % unit
% \SetGraphUnit{1.4} 
% % styles
% \GraphInit[vstyle=Normal] 
% \SetVertexNormal[Shape=circle,MinSize=0.5cm,LineWidth =1pt]
% \tikzset{VertexStyle/.append style = {font=\Large\bfseries},thick} 
% % vertices  
% \Vertex{1} 
% \SOWE(1){2} 
% \SOEA(1){3} 
% \SOEA(2){4} 

% % intern edges
% % \Edges(1,3) 
% % loops
% % \Loop[dist=3cm,dir=NO,style={thick},label=$0.1$,labelstyle=above](1)  
% % \Loop[dist=3cm,dir=WE,style={thick},label=$0.4$,labelstyle=left](2)  
% % \Loop[dist=3cm,dir=EA,style={thick},label=$0.6$,labelstyle=right](4)
% % intern labels 
% % \path[every node/.style={swap,auto}]    (1) to node {0.2} (2)
% %                                             to node {0.3} (3)
% %                                             % to node {0.8} (2)
% %                                             to node {0.4} (1); 
% % draw extern edges and label
% \draw[<-] (1) to [bend right] node [above right] {$p_{21}$} (2);

% \draw[<-] (2) to [bend right] node [below left]  {$p_{42}$} (4);  
% \draw[<-] (4) to [bend right] node [above left]  {$p_{14}$} (1);  
% \draw[<-] (1) to [bend left] node [below right] {$p_{31}$} (3);

% \end{tikzpicture} 

%     \caption{Propositional variables in cyclic graphs}
%     \label{fig:sample_cycle}
% \end{figure}

% \subsection{Propositional Variables and MAXSAT problem}
 First, we introduce propositional logic and the definition of a MAXSAT problem for cycle resolving. For a directed graph $G =\langle E, V \rangle$, if there is an edge from a node $v$ to $w$ ($v, w \in V$), then we have $(v, w) \in E$. A \textit{propositional variable}  $p_{v,w}$ represents if there is a directed edge from $v$ to $w$. If False then we remove the edge. An assignment is to associate all the variables with True or False. 
 
 In Figure~\ref{fig:exampegraph}, we can find the following five nodes in the blue cycle: $v_1$, $v_2$, $v_3$, $v_7$ and $v_8$. To resolve it, we need to remove at least one edge. Hence, we set at least one of the following variables $p_{1,2}$, $p_{2,3}$, $p_{3,7}$, $p_{7,8}$, $p_{8,1}$ to False. Equivalently, we need to make the \textit{clause} $s = \neg p_{1,2}\lor \neg p_{2,3} \lor \neg p_{3,7} \lor \neg p_{7,8} \lor \neg p_{8,1} $ evaluate to True. 
 
In this manner, for each cycle $c_i \in SC$, we generate a clause $s_i$. To resolve all the cycles, we simply need to find an assignment so that all the clauses evaluate to True. It is easy to notice that an easy solution is to remove all the edges (i.e. simply set all propositional variables to false). This amounts to removing all edges and thus removing all cycles. To maintain as much information as possible, we formulate this problem as a Partially Weighted MAXSAT problem. 
 
 A \emph{Partially Weighted MAXSAT} problem has constraints in two forms: \textit{soft constraints} and \textit{hard constraints}. The fact that we need to resolve all the cyclic connections corresponds to  hard constraints (as encoded above). A soft constraint is in the same form except a weight $w_i$ associated with each clause, denoted $(c_i, w_i)$. The goal of this type of problems is to satisfy all the hard constraints while maximising the sum of the weights associated with the satisfied soft constraints, and thus it is a constrained optimisation problem. 
 
 In this case, our soft constraints are simply each of the propositional variables corresponding to the edges (relations). For fairness, we assign a fixed identical weight each. In this way, the MAXSAT procedure will remove all cycles (i.e. satisfying the hard constraints), while keeping as many of the relations as possible (i.e. maximally satisfying the soft constraints). For simplicity, we use in the term MAXSAT in reference to Partially Weighted MAXSAT. For the example above, the hard constraint is $s$ as encoded above and the corresponding soft constraints are propositional variables with identical weights of 1 each.

 The problem of weighted MAXSAT is $\Delta^P_2$-complete (a linear number of calls on instance size to a SAT oracle). When weights are equal, it is $\Delta^{P_2}[\log n]$-complete (a logarithmic number of calls to a SAT oracle). Our algorithm takes all the weights identically and the number we call the MAXSAT solver $i$ is bounded by the number of nodes $n$. To handle combinatorial explosion, we introduce a bound $B$ to limit the number of simple cycles at each iteration. Although both cycle-finding and cycle-resolving  in this setting are intractable, efficient solvers/programs exist that can solve problems of realistic size. 

% https://es-static.fbk.eu/events/satsmtschool12/slides/1x04_SS12.pdf

% combinatorial explosion 

% \subsection{First attempt} 

% My first attempt is to first form this problem to a MAXSAT problem. We need two sets of constraints: the hard constraints and the soft constraints. The hard ones are those that we have to satisfy. In other words, we would like to resolve the cyclic subclass relations. The soft constraints are to maximise the amount of categorical information of classes. 

% For memory efficiency, we first export these triples so we do not have to load the LOD-a-lot in memory for the 

% What is basis cycles. 
% How is it related to the 

% Hypothesis 1: By resolving the cyclic subclass relations using the \textbf{basis cycles}, we can resolve a significant amount cyclic relations. 

% Encode: 

\subsection{Implementation}
\label{implementation}
The SUBMASSIVE system was implemented in Python and takes advantage of both the \texttt{networkx} Python package\footnote{\url{https://networkx.github.io/}} and the Python binding of Z3 SMT solver,\footnote{\url{https://github.com/Z3Prover/z3}} together with interaction with the LOD-a-lot\footnote{\url{http://lod-a-lot.lod.labs.vu.nl/}} HDT file using the PyHDT library.\footnote{\url{https://github.com/Callidon/pyHDT}}  % describe the scale of our data
% Despite other large knowledge graphs, the results of this paper is solely about the LOD-a-lot knowledge graph \cite{lod}. 
The LOD-a-lot dataset represents the graph merge of 650K datasets crawled from the LOD Cloud in 2015 \cite{lod}. It contains over 28 billion triples, making it one of the largest publicly available knowledge graphs. The corresponding subgraph $G_\text{cat}$ has over 4 million edges. There are several cycles nested inside each other, which results in combinatorial explosion. 
% It turned out to have many cycles nested inside each other. As a result, it was impossible to retrieve all the cycles due to combinatorial explosion. 

Such a complex integration resulted in lack of efficiency in memory use. We therefore wrote a separate script to obtain the edges removed in comparison against the equivalent classes and the sameAs.cc dataset.\footnote{We use the dataset (closure\_099) published at \url{https://zenodo.org/record/3345674}} After comparison, we identified 755 relations as unnecessary and removed them. These relations together with its decisions are loaded to the system.

At each iteration, the \texttt{find\_cycle} method uses an optional list of classes as source $S$ and performs a depth-first search. If no cycles are found starting from $S$, a class is arbitrarily chosen and repeatedly searched until there was a cycle or ended up with an exception of no cycle. To avoid retrieving the same cycle, we removed one random relation of the cycle retrieved from a copy of $G_\text{cat}$ and continue to the next iteration with this new subgraph $G'_\text{cat}$. 

Our simple cycles are retrieved using the \texttt{simple\_cycle} method. The algorithm uses a trick to block duplicated search from a root and then build the elementary path from the searching results. Details are in \cite{simple}.

It is worth noting that a rare case is that only one cycle was found whose size exceeded the soft size bound $B$. In such case, a random relation would be removed from this cycle. An update of the algorithm was to limit the minimum amount of cycles to 3 (unless no more cycles found). 

Using SUBMASSIVE, a user can retrieve, for any given class, all its super-classes (i.e. its transitive closure) without worries of cyclic cases. Using a variant of the system, we resolve the cycles in the graph of \texttt{rdfs:subPropertyOf} and publish the resulting hierarchy at the same place in N-Triples format. The size of the resulting cycle-free datasets is 829.6MB for subClassOf, and 14.6MB for subPropertyOf.

\section{Evaluation}
\label{eva}

We present our evaluation of the fully automated version (Section \ref{fa}) and then compare it against a semi-automatic version (Section \ref{semi}).

\subsection{Evaluation of the Fully Automated Version}
\label{fa}
\begin{figure}
\begin{center}
	\includegraphics[scale=0.55]{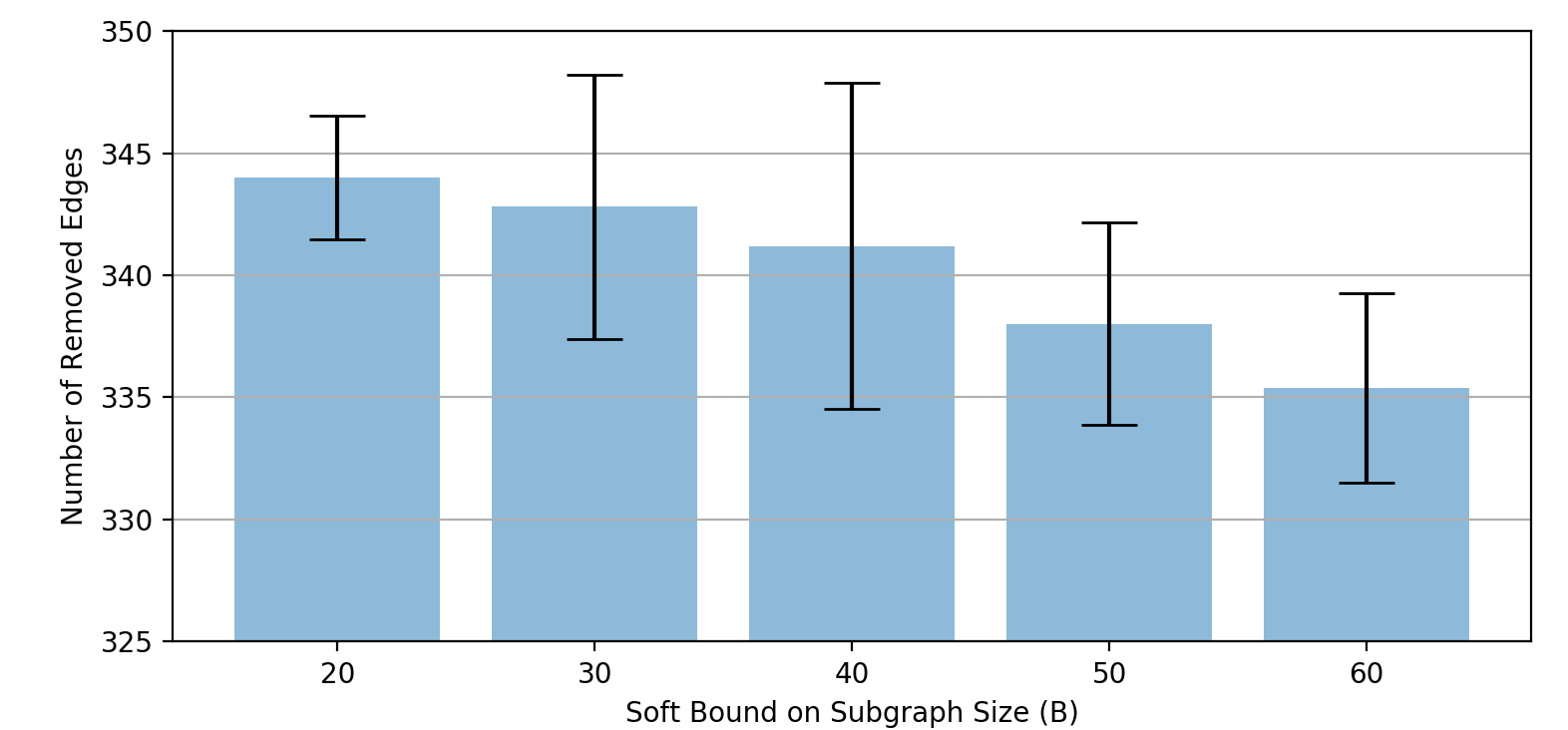}
	\captionof{figure}{Number of removed edges against the soft bound on size of subgraph at each iteration (error-bars represent the standard deviation).}
\label{fig:evaremoved}
\end{center}
\end{figure}

 Figure~\ref{fig:evaremoved} illustrates how the total amount of removed relations decreases as we increase the size $B$. The blue bars indicates the average and the error-bars represent the standard deviation. The best run from our experiment has 330 relations were removed as a result (soft bound $B = 60$). The algorithm is not deterministic because we remove a random edge for each cycle when retrieving the neighbourhood. For each value of $B$ between 20 and 60, we performed the experiment 5 times to obtain the mean and variance (since the algorithm is not deterministic).\footnote{All our experiments were conducted on the following machine: Intel Xeon CPUs (E5-2630 v3 @ 2.40GHz) with a RAM of 264GB. The processing time varies between 26 and 99 minutes. We set our time limit to 2 hours, which limits the size of subgraph $N$ to 70.} When $B \geq 70$, the system faced combinatorial explosions for some runs: there can be over 1 million simple cycles and one iteration may take over one hour or even longer. The results are therefore not included. The variation (as illustrated in Figure~\ref{fig:evaremoved} as error bars) of the removed relations is due to the random removal of edges when obtaining subgraphs. The figure shows a decrease in number of edges removed as the bound enlarges.

\begin{figure}[!ht]
\centering
\begin{subfigure}[b]{0.95\textwidth}
   \includegraphics[width=1\linewidth]{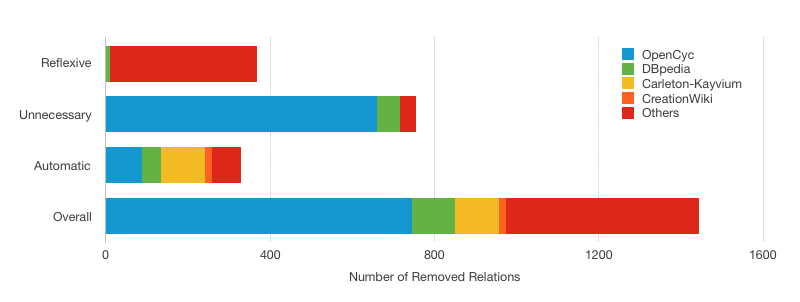}
   \caption{Fully automatic (FA)}
   \label{fig:automatic_final} 
\end{subfigure}

\begin{subfigure}[b]{0.95\textwidth}
   \includegraphics[width=1\linewidth]{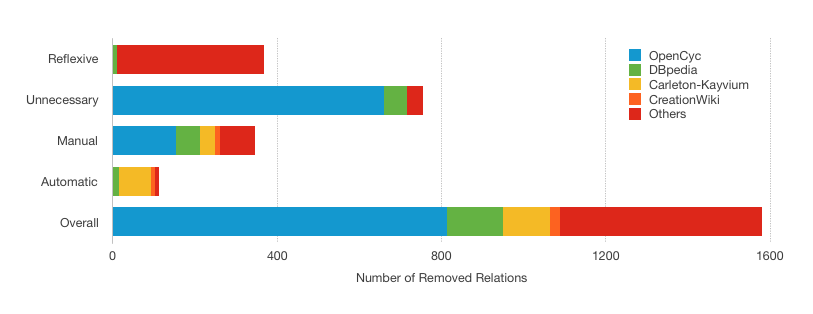}
   \caption{Semi-automatic (SA)}
   \label{fig:theBarManual}
\end{subfigure}

\caption[Two numerical solutions]{An analysis of removed relations in stacked bar-chart.}
\end{figure}

 Figure \ref{fig:automatic_final} shows the sources of the removed relations. After comparison against the equivalent classes and the sameAs.cc dataset, we identified 38 and 717 relations respectively, summing up to a total of 755 relations. A closer examination unveils that 87.2\% of the unnecessary relations have a subject with the OpenCyc namespace. It is clear that such cycles were mostly inherited from the OpenCyc, DBpedia and a knowledge base by Kayvium.\footnote{We use carleton-kayvium as an abbreviation for \url{http://http-server.carleton.ca/~rgarigue/ontologies/www.kayvium.com/Regional_registry}. The knowledge graph suffers heavily from erroneous subclass relations.} The source of error could trace back to when these knowledge graphs were constructed from structured/semi-structured knowledge or extracted from text \cite{refinement}. In addition, a few interdisciplinary relations were removed (less than 12, the exact number depends on the run). 
%  It worth mentioning that there were some very densely related classes in the carleton-kayvium knowledge base and a significant amount of iteration is about resolving its cycles and the relations are mostly erroneous.   
 
\subsection{Evaluation against the Semi-automated Version}
\label{semi}
Since there is no gold standard available, we manually checked the correctness of relations resulting in size-two cycles. There were a total of 229 size-two cycles, forming 458 relations. After the manual pre-processing,\footnote{We relied on the IRI, labels, comments and other information online for decision making. This manual processing removed relations more strictly than the automatic processing which solely aims at resolving cycles. If $A \equiv B$, we consider both relations \textit{unnecessary} and remove them. If unknown, both relations remain.} there were in total 345 relations identified as erroneous to the best of our knowledge (Figure \ref{fig:theBarManual}). Following that, we conduct the automatic part. The best run has 131 relations removed ($B$ = 60).

We compare the removed relations in two approaches: fully automatic (FA) v.s. semi-automatic (SA). Recall that there were 330 relations removed in FA against 345 relations removed in the manual processing and 131 automatic processing of SA. Out of the 330 relations removed during FA, 165 of them were also removed in the manual process of SA (50.0\%) and 57 of them (17.27\%) in the automatic process of SA. These two sets summed up to 222 triples (67.27\%). A reason that the precision was not higher is that the weights on relations for size-two cycles were identical when resolving automatically. Thus, there was an equal chance for a relation to be removed unless it was nested in other cycles in the iteration.  Moreover, no entry from the FA results were to be removed in the SA process, which indicates that our result has a low false-positive rate. 
% This result confirmed that the results of automatic process was considerably reliable. 
% In addition, out of the 330 relations removed, 28 appeared to be labelled as `unknown' in the manual decision process. 

% \begin{table}[]
% \centering
% \begin{tabular}{|l|l|l|l|l|}
% \hline
% Experiment        & Name & Content & Triple & Size \\ \hline
%   Original Data     &   subC-all   &  all the subclass subsumption relations       &    4,461,717    &   XMB    \\ \hline
% \multirow{2}{*}{Refined-FA} & pre-auto-FA     &    relations removed in pre-processing and automatic process     &        &      \\ \cline{2-5} 
%                   &   refined-subC-FA  &    remain relations after processing     &        &      \\ \hline
% \multirow{3}{*}{Refined-SA} &   pre-FA   &    pre-processing with manual decisions     &        &      \\ \cline{2-5} 
%                   &   auto-SA   &      relations removed in automatic process   &        &      \\ \cline{2-5} 
%                   &   refined-subC-SA   &     relations remain after SA processing    &        &      \\ \hline
% \multirow{3}{*}{SubProperty} &  subP-all    &    all the subproperty subsumption relations      &    80,356    &      \\ \cline{2-5} 
%                   &   pre-subP   &     pre-processing with manual decisions     &    1,957     &     \\ \cline{2-5} 
%                   &   refined-subP   &     relations remain    &     78,399   &      \\ \hline
% \end{tabular}
% \caption{Datasets}
% \label{tab:datasets2}
% \end{table}

\section{Discussion}

\label{discuss}
We have presented an algorithm for the refinement of subclass relations in very large knowledge graphs and also compared against its semi-automatic version. A bottleneck of our algorithm is that we enumerate all simple cycles for encoding. The subgraph $G_\text{cat}$ of LOD-a-lot turns out to have many cycles nested inside each other. Due to combinatorial explosion, we are only able to perform the algorithm less than one million simple cycles (timeout otherwise). This ensures the subgraph $G[N]$ at the iteration to be cycle-free but limits the size of subgraph $N$ to 60. This implies that the amount of edges we remove is not global minimal but an approximation to it. Another alternative approach would be to retrieve all the nodes involved in cycles and then compute a limited amount of simple cycles of the corresponding subgraph at each iteration until cycle-free. 

LOD-a-lot was built from a crawl of the LOD Laundromat in 2015 \cite{lod}. In the most recent version of DBpedia (2019), there is no cycle found. In contrast to other removed relations, those from the creationwiki (\url{http://creationwiki.org}) come with a biblical worldview. Despite its cyclic assertions, we did not find any contradictions that were directly due to this difference in worldview. This highlights that in LOD-a-lot, multiple contradicting worldviews may be present.

Although there are cycles involving cross-domain relations, most removed relations are within the same domain. An explanation is that since relations within the domain are also involved in other cycles, for the sake of optimisation, they are removed. Another possible reason is that our approach prioritises the resolving of local cycles. Most removed ones are between domains \url{http://www.daml.org}, \url{http://ontology.ihmc.us}, and \url{http://www.w3.org}. 

% limit of evaluation
Considering there is no ground truth in very large knowledge graphs, our evaluation has its limits. An alternative way to perform evaluation could be to add or remove some subclass relations from a verified knowledge graph and then use our methods to identify them. 

% Since a large knowledge graph may integrate various data from different resources, it worth mentioning the tolerance of difference. Therefore, our evaluate is not absolute. 

\section{Conclusion and Future Work}

 We have presented an algorithm for the refinement of subclass relations in very large knowledge graphs. Our algorithm takes a hybrid approach, combining the use of network/graph theory and automated reasoning. The fully automatic version does not use external knowledge but still generates relatively reliable results. With some additional manual processing we show that these results can be considerably improved without sacrificing scalability. We have tested our approach on the LOD-a-lot dataset. The resulting cycle-free subclass hierarchy is available on Zenodo with the DOI: 10.5281/zenodo.3693802.

% During the pre-processing as presented in Section \ref{pre}, we noticed that unnecessary connections came mostly from the OpenCyc knowledge graph (87.2\%), which indicates that it has some redundancy. 
While this work considers a class subsumption relation as unnecessary when there also exists an \texttt{owl:equivalentClass} relation, a future work could be to replace the \texttt{owl:equivalentClass} relations by two subsumption relations. Another possible future direction is to convert all the class concepts in \texttt{owl:sameAs} relation to one uniform class node while maintaining the relations with other classes. There could be more relations worth removing. 

% The result of this project, together with the refined owl:sameAs dataset provide suggestions on the refinement of the LOD-a-lot. The system was designed to return refined results of queries to HDT taking consideration of such refined relations. 

Our work also examines the quality of knowledge graphs. For example, the carleton-kayviu and creationwiki knowledge graphs have complex subclass cycles in their ontology while OpenCyc needs a revisit on the use of subclass and equivalent class relations. 

Finally, with this cycle-free graph, we are now able to obtain a reliable subclass transitive closure of an entity. This provides the data to perform Machine Learning on large-scale, such as evaluating the accuracy of transitive relations regarding different knowledge graph embeddings. In other words, if an entity $e$ is of type $A$, and $A$ is a subclass of $B$, we would like to know how likely $e$ is also of type $B$ (or its super classes, without having to deal with incorrect cycles). 

\paragraph{Acknowledgements} The project is a part of the MaestroGraph project funded by the NWO TOP grant. We thank Luigi Asprino, Wouter Beek and Jacopo Urbani and anonymous reviewers for their helpful advice.

\bibliographystyle{unsrt}
\bibliography{main}

\begin{thebibliography}{10}

\bibitem{refinement}
Heiko Paulheim.
\newblock Knowledge graph refinement: A survey of approaches and evaluation methods.
\newblock {\em Semantic Web}, 8(3):489--508, 2017.

\bibitem{ma2014learning}
Yanfang Ma, Huan Gao, Tianxing Wu, and Guilin Qi.
\newblock Learning disjointness axioms with association rule mining and its application to inconsistency detection of linked data.
\newblock In {\em Chinese Semantic Web and Web Science Conference}, pages 29--41. Springer, 2014.

\bibitem{paulheim2014improving}
Heiko Paulheim and Christian Bizer.
\newblock Improving the quality of linked data using statistical distributions.
\newblock {\em International Journal on Semantic Web and Information Systems (IJSWIS)}, 10(2):63--86, 2014.

\bibitem{paulheim2014identifying}
Heiko Paulheim.
\newblock Identifying wrong links between datasets by multi-dimensional outlier detection.
\newblock In {\em WoDOOM}, pages 27--38, 2014.

\bibitem{cuzzola2015filtering}
John Cuzzola, Ebrahim Bagheri, and Jelena Jovanovic.
\newblock Filtering inaccurate entity co-references on the linked open data.
\newblock In {\em International DEXA Conference}, pages 128--143. Springer, 2015.

\bibitem{hogan2012scalable}
Aidan Hogan, Antoine Zimmermann, J{\"u}rgen Umbrich, Axel Polleres, and Stefan Decker.
\newblock Scalable and distributed methods for entity matching, consolidation and disambiguation over linked data corpora.
\newblock {\em Web Semantics: Science, Services and Agents on the World Wide Web}, 10:76--110, 2012.

\bibitem{papaleo2014logical}
Laura Papaleo, Nathalie Pernelle, Fatiha Sa{\"\i}s, and Cyril Dumont.
\newblock Logical detection of invalid sameas statements in rdf data.
\newblock In {\em International Conference EKAW}, pages 373--384. Springer, 2014.

\bibitem{gueret2012assessing}
Christophe Gu{\'e}ret, Paul Groth, Claus Stadler, and Jens Lehmann.
\newblock Assessing linked data mappings using network measures.
\newblock In {\em Extended Semantic Web Conference}, pages 87--102. Springer, 2012.

\bibitem{raad2018detecting}
Joe Raad, Wouter Beek, Frank van Harmelen, Nathalie Pernelle, and Fatiha Sa{\"\i}s.
\newblock Detecting erroneous identity links on the web using network metrics.
\newblock In {\em International Semantic Web Conference}, pages 391--407. Springer, 2018.

\bibitem{simple}
Donald~B Johnson.
\newblock Finding all the elementary circuits of a directed graph.
\newblock {\em SIAM Journal on Computing}, 4(1):77--84, 1975.

\bibitem{lod}
Javier~David Fernandez~Garcia, Wouter Beek, Miguel~A Mart{\'\i}nez-Prieto, and Mario Arias.
\newblock \text{LOD-a-lot}: A queryable dump of the lod cloud.
\newblock 2017.

\end{thebibliography}

\end{document}